\documentclass[preprint,12pt]{cas-sc}
\usepackage[authoryear]{natbib}
\usepackage{lineno}
\definecolor{grey}{rgb}{0.9,0.9,0.9}
\usepackage{textgreek}

\def\tsc#1{\csdef{#1}{\textsc{\lowercase{#1}}\xspace}}
\tsc{WGM}
\tsc{QE}
\begin{document}
\let\WriteBookmarks\relax
\def\floatpagepagefraction{1}
\def\textpagefraction{.001}

\shorttitle{Io Atmospheric Variability}    

\shortauthors{Giles et al.}  

\title [mode = title]{Seasonal and longitudinal variability in Io's SO\textsubscript{2} atmosphere from 22 years of IRTF/TEXES observations}  

% Authors
\author[1]{Rohini S. Giles}[type=editor,
                        auid=000,bioid=1,
                        orcid=0000-0002-7665-6562]
\cormark[1]
\ead{rohini.giles@swri.org}

\author[2]{John R. Spencer}
\author[2,3]{Constantine C. C. Tsang}
\author[1]{Thomas K. Greathouse}
\author[4]{Emmanuel Lellouch}
\author[5]{Miguel A. López-Valverde}

\affiliation[1]{organization={Space Science Division, Southwest Research Institute},
            city={San Antonio},
            state={Texas},
            country={USA}}
\affiliation[2]{organization={Planetary Science Division, Southwest Research Institute},
            city={Colorado},
            state={Texas},
            country={USA}}
\affiliation[3]{organization={Khan Enterprises LLC},
            city={Boulder},
            state={Colorado},
            country={USA}}
\affiliation[4]{organization={LEISA, Observatoire de Paris/PSL},
            city={Meudon},
            country={France}}   
\affiliation[5]{organization={Instituto de Astrofisica de Andalucia},
            city={Granada},
            country={Spain}}  

\cortext[1]{Corresponding author}


\begin{abstract}
Between 2001 and 2023, we obtained high spectral resolution mid-infrared observations of Io using the TEXES instrument at NASA's Infrared Telescope Facility. These observations were centered at 529.8 cm\textsuperscript{-1} (18.88 \textmu m) and include several SO\textsubscript{2} absorption lines. By modeling the shapes and strengths of these absorption lines, we are able to determine how Io's SO\textsubscript{2} atmospheric density varies over the 22-year time period, covering nearly two Jovian years. Previous analysis has shown that the density of Io's atmosphere on the anti-Jovian hemisphere exhibits clear seasonal temporal variability, which can be modeled as the sum of a seasonally-varying frost sublimation component and a constant component, assumed to be volcanic. The new data show that the seasonal pattern repeats during the second Jovian year, confirming the importance of sublimation support. The considerable longitudinal variability in Io's atmospheric density found in previous work is also stable over the second Jovian year with the SO\textsubscript{2} column density on the Jupiter-facing hemisphere being 5--8 times lower than the anti-Jovian hemisphere. For the first time, we detect seasonal variability on the Jupiter-facing hemisphere as well.  This can also be modeled as a combination of sublimation and a small constant source. The lower atmospheric density on the Jupiter-facing hemisphere can plausibly be explained by the daily Jupiter eclipses, which decrease the surface temperature and therefore reduce the sublimation-driven component of the atmosphere, combined with a lower level of volcanic activity directly emitting SO\textsubscript{2} into the atmosphere.
\end{abstract}


\begin{highlights}
\item Io's SO\textsubscript{2} atmospheric density varies seasonally according to heliocentric distance
\item The atmosphere is supported by sublimation of surface SO\textsubscript{2} frost on both hemispheres 
\item The atmospheric density is consistently much lower on the Jupiter-facing hemisphere
\item Eclipses and inhomogeneous volcanic emissions may drive the longitudinal variability
\end{highlights}

% Keywords
\begin{keywords}
Io \sep
Atmospheres, structure \sep 
Infrared observations \sep
Spectroscopy
\end{keywords}

\maketitle

% Main text

%\linenumbers 

\section{Introduction}
\label{sec:introduction}

Io's SO\textsubscript{2}-dominated atmosphere is thought to be supported by a combination of vapor pressure equilibrium with surface SO\textsubscript{2} frost and direct supply from volcanic plumes, making it one of the most unusual atmospheres in the solar system~\citep{depater23}. While the evidence for significant support by frost sublimation is now very strong, as described below, the relative roles of sublimation and volcanism are not yet fully understood. Several types of observations of Io's atmosphere have previously been used to tease apart the contributions from the two sources, primarily taking advantage of the fact that frost sublimation depends on the amount of sunlight incident on the surface but volcanic activity does not. These studies have considered the geographical variations in the atmospheric density~\citep{jessup04,spencer05,lellouch15}, the auroral and atmospheric response to eclipses~\citep{saur04,retherford07,tsang16,depater20}, the atmospheric density as a function of local time~\citep{jessup15,lellouch15}, and the seasonal variability over the course of a Jovian year~\citep{tsang12,tsang13b}. In this section, we summarize the results from these previous studies and introduce our contribution to the topic.

Studies using both ultraviolet and infrared observations have found that the SO\textsubscript{2} atmospheric density peaks on the anti-Jovian hemisphere of Io, with column densities approximately 10 times higher than at Jupiter-facing longitudes~\citep{jessup04,spencer05,tsang13}. This broad geographical distribution correlates with both the SO\textsubscript{2} frost distribution~\citep{mcewen88} and the locations of active plumes~\citep{mcewen83,spencer07,geissler08}, and could therefore support either frost sublimation or volcanism as the primary atmospheric support mechanism. However, spatially-resolved observations from~\cite{jessup04} and~\cite{lellouch15} have found that the increase in SO\textsubscript{2} over active plumes is relatively modest, suggesting that frost sublimation dominates, with volcanism as a smaller component.

Io's auroral emissions and SO\textsubscript{2} column densities have both been observed to decrease significantly during eclipse by Jupiter. \cite{saur04} and \cite{retherford07} both studied the response of Io's ultraviolet auroral emissions to eclipse and found that the decrease in emissions must be due to a substantial decrease in the atmospheric density. Both studies inferred that frost sublimation must be responsible for at least 90\% of the atmosphere when it is illuminated by the sun, but that volcanism may dominate during eclipses and on the nightside. \cite{tsang16} used mid-infrared observations to show that Io's SO\textsubscript{2} atmosphere collapses by a factor of 5 as the moon moves into Jupiter's shadow, concluding that the Jupiter-facing atmosphere is sublimation-dominated. \cite{depater20} measured Io in the 1-mm band as it moved into and out of eclipse and observed the SO\textsubscript{2} atmosphere partially collapse and reform. While sublimation must play a significant role in supporting the atmosphere, they find that the level of collapse suggests that volcanic sources contribute 30--50\%.

Diurnal observations of Io's atmosphere have produced somewhat conflicting results. \cite{lellouch15} measured moderate diurnal variations in the SO\textsubscript{2} column density across the disk of Io using spatially-resolved infrared observations. This observation provides evidence for an atmosphere that is at least partially sublimation-driven. In contrast, \cite{jessup15} used near-ultraviolet spectroscopy from the Space Telescope Imaging Spectrograph (STIS) on the Hubble Space Telescope (HST) to compare atmospheric densities obtained at the same longitude but different local time and found negligible differences. This lack of variability is consistent with primarily volcanic support, though these observations could also be explained by high thermal inertia of the sublimating frost which would reduce the amplitude of diurnal temperature and vapor pressure variations.

SO\textsubscript{2} absorption lines at 530 cm\textsuperscript{-1} (19 µm) were first measured in Io's atmosphere in 2001 using TEXES, a high-resolution mid-infrared spectrograph at NASA's Infrared Telescope Facility (IRTF)~\citep{spencer05}. Since then, regular observations of Io's atmosphere have continued to be made in order to track the seasonal variability.~\cite{tsang12,tsang13b} tracked the variation of the peak anti-Jovian density from 2001 to 2013 and found a large increase in SO\textsubscript{2} atmospheric density near perihelion, when surface frost is expected to be warmest and its vapor pressure highest, indicating a major role for frost sublimation support on this hemisphere.  However their models of seasonal variations in frost temperature and vapor pressure indicated that an additional constant component was also required to match the observed amplitude of the seasonal variability. This constant component was presumed to be volcanic; while volcanic activity has been observed to vary on timescales of months or less~\citep[e.g][]{dekleer17,tate23}, it does not have any seasonal variability and could therefore plausibly be considered a constant source over longer timescales. 

In this paper, we extend the previous TEXES/IRTF observations from~\cite{spencer05} and~\cite{tsang12,tsang13b} and present a time series of Io's atmospheric SO\textsubscript{2} density covering 2001-2023. This 22-year dataset covers almost two Jovian years and constitutes one of the
longest continuous series of observations of Io’s atmosphere using the same instrument and telescope, along with Ly-\textalpha~imaging from HST/STIS. However,~\cite{giono21} reanalyzed observations from~\cite{feaga09} and found that the Ly-\textalpha~images are not sensitive enough to the SO\textsubscript{2} density to track seasonal changes, making the IRTF/TEXES observations uniquely valuable. The long-term observations allow us to confirm that the previously observed seasonal behaviour continues, and look for any interannual variability. Recent observations have been obtained at a higher cadence than previously, and have covered a greater range of longitudes, allowing us to better constrain the seasonal model and to search for any seasonal atmospheric variability on the less dense Jupiter-facing hemisphere. In Section~\ref{sec:observations} we describe the observations, the data reduction and the spectral modeling used to retrieve the SO\textsubscript{2} density from each observation. The longitudinal variability, and the temporal variability on both the anti-Jovian and Jupiter-facing hemispheres are presented in Section~\ref{sec:results} and these results are discussed in Section~\ref{sec:discussion}.

\section{Observations}
\label{sec:observations}

\subsection{TEXES observations}

Between 2001 and 2023, regular observations of Io were obtained using the TEXES instrument~\citep{lacy02}, a high-resolution mid-infrared spectrograph that is mounted at the 3-m IRTF telescope on Mauna Kea. Data were obtained on 24 observing runs, summarized in Table~\ref{tab:run_summary}. The first 10 of these observing runs (2001--2013) were previously analyzed in~\cite{spencer05} and~\cite{tsang12,tsang13b}. The entire dataset now spans almost two Jovian years, allowing seasonal trends to be clearly observed and any inter-annual variability to be investigated.

\begin{table}[hbt!]
\begin{center}
\begin{tabular}{|>{\centering\arraybackslash}m{1.8cm} |>{\centering\arraybackslash}m{1.8cm} | >{\centering\arraybackslash}m{2.0cm} | >{\centering\arraybackslash}m{6.6cm} | >{\centering\arraybackslash}m{2.0cm}|}% >{\centering\arraybackslash}m{1.75cm} |} 

 \hline

Start date & End date & Number of observations & Io central longitudes ($^{\circ}$W) & Heliocentric distance (AU) \\%& Heliocentric longitude ($^{\circ}$)\\

 \hline

2001-11-16 & 2001-11-18 & 3 & 16, 38, 213 & 5.15 \\%& 96.8 \\
\rowcolor{grey} 2002-12-06 & 2002-12-19 & 7 & 23, 33, 44, 93, 196, 214, 321 & 5.30 \\%& 129.0 \\
2004-01-06 & 2004-01-08 & 3 & 99, 139, 153 & 5.41 \\%& 159.5 \\
\rowcolor{grey} 2005-01-16 & 2005-01-19 & 4 & 104, 156, 268, 306 & 5.45 \\%& 188.2 \\
2007-04-14 & 2007-04-18 & 6 & 68, 78, 165, 214, 323, 331 & 5.34 \\%& 250.8 \\
\rowcolor{grey} 2009-06-02 & 2009-06-07 & 9 & 107, 116, 124, 156, 218, 225, 265, 273, 282 & 5.06 \\%& 315.7 \\
2010-06-02 & 2010-06-03 & 3 & 158, 164, 317 & 4.97 \\%& 348.2 \\
\rowcolor{grey} 2012-01-08 & 2012-01-17 & 10 & 90, 98, 106, 111, 129, 133, 137, 145, 152, 311 & 4.97 \\%& 41.8 \\
2012-10-07 & 2012-10-11 & 8 & 43, 50, 210, 221, 231, 294, 303, 314 & 5.04 \\%& 66.2 \\
\rowcolor{grey} 2013-02-04 & 2013-02-12 & 12 & 103, 110, 124, 133, 144, 153, 237, 245, 283, 293, 325, 336 & 5.08 \\%& 77.1 \\
2014-02-20 & 2014-02-20 & 1 & 203 & 5.21 \\%& 109.0 \\
\rowcolor{grey} 2014-12-07 & 2014-12-12 & 4 & 157, 165, 201, 210 & 5.31 \\%& 132.8 \\
2015-11-15 & 2015-11-15 & 2& 138, 148 & 5.41 \\%& 159.3 \\
\rowcolor{grey} 2016-04-30 & 2016-04-30 & 2 & 198, 210 & 5.44 \\%& 172.1 \\
2018-02-10 & 2018-02-10 & 1 & 203 & 5.43 \\%& 221.4 \\
\rowcolor{grey} 2018-07-19 & 2018-07-19 & 1 & 98 & 5.39 \\%& 233.6 \\
2019-04-15 & 2019-04-23 & 8 & 59, 154, 213, 221, 260, 269, 324, 331 & 5.32 \\%& 254.9 \\
\rowcolor{grey} 2019-08-15 & 2019-08-20 & 4 & 224, 272, 281, 290 & 5.28 \\%& 264.6 \\
2021-07-04 & 2021-07-07 & 12 & 56, 63, 79, 103, 262, 270, 277, 283, 310, 318, 326, 332 & 5.04 \\%& 322.9 \\
\rowcolor{grey} 2021-09-27 & 2021-09-29 & 3 & 52, 98, 106 & 5.01 \\%& 330.4 \\
2021-11-06 & 2021-11-09 & 8 & 74, 83, 91, 97, 122, 130, 234, 243 & 5.00 \\%& 334.0 \\
\rowcolor{grey} 2022-06-26 & 2022-07-03 & 13 & 100, 108, 200, 209, 218, 248, 256, 297, 304, 312, 313, 320, 330 & 4.96 \\%& 355.4 \\
2022-09-15 & 2022-09-18 & 12 & 73, 80, 88, 114, 122, 130, 230, 238, 246, 265, 273, 280 & 4.96 \\%& 2.4 \\
\rowcolor{grey} 2023-07-20 & 2023-07-25 & 14 & 75, 83, 118, 126, 133, 165, 232, 242, 277, 285, 291, 321, 329, 336 & 4.96 \\%& 30.8 \\

 \hline

\end{tabular}
\caption{Summary of IRTF observing runs contributing to this study: the start and end dates of each run, the number of independent observations of Io obtained over the course of the observing run, the central longitudes of each of these observations and the heliocentric distance of the Jovian system.}
\label{tab:run_summary}
\end{center}
\end{table}

\begin{figure*}
\centering
\includegraphics[width=14cm]{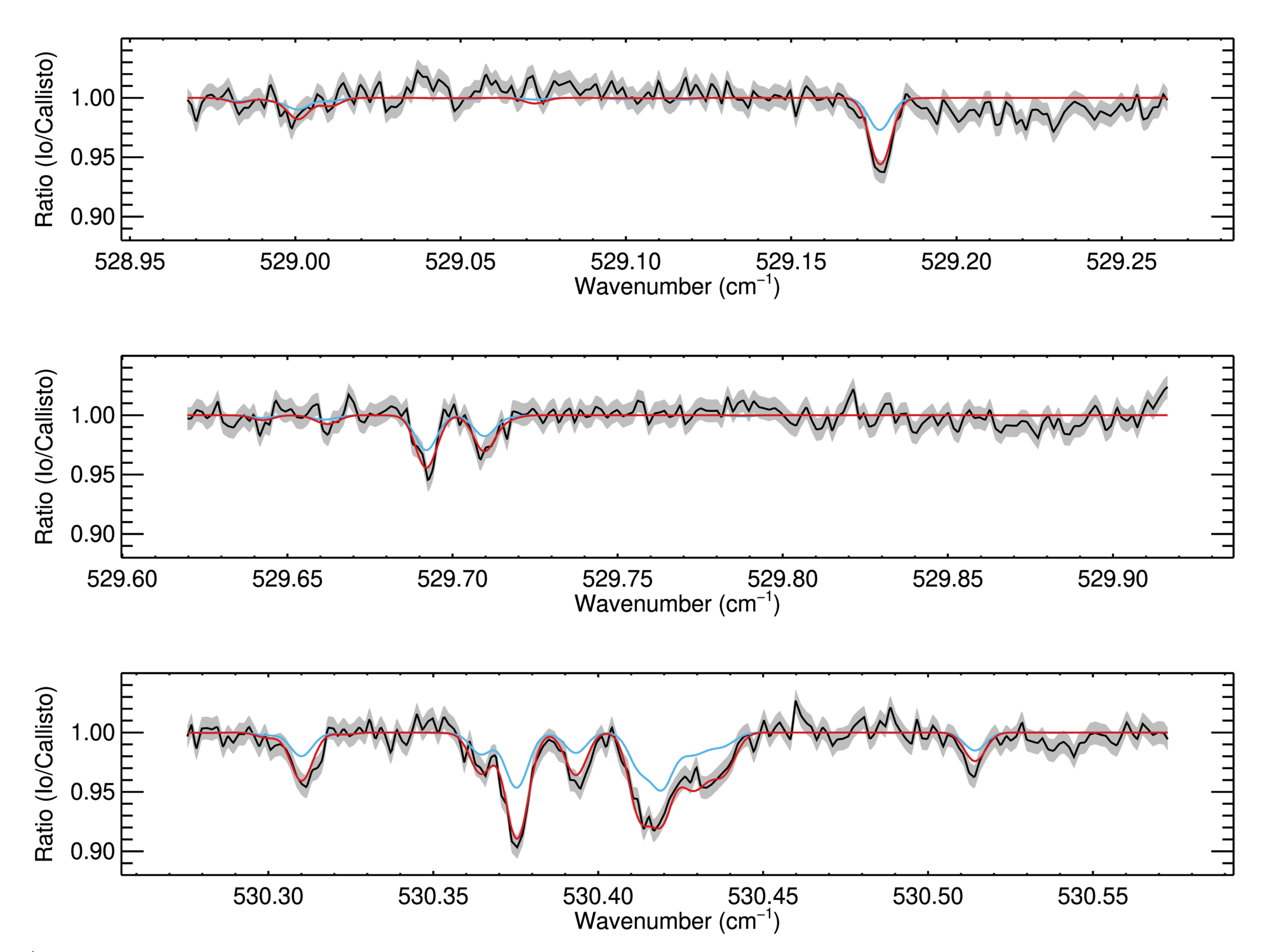}
\caption{Black: An example reduced Io spectrum, obtained on 29 June 2022 at an Io central longitude of 200$^{\circ}$W. The noise on the spectrum is shown in gray. The spectrum has been divided by a Callisto spectrum, flattened and normalized. Absorption features are due to SO\textsubscript{2}. Red: The corresponding best-fit model spectrum obtained from the spectral modeling technique, assuming an atmospheric temperature of 108 K. Blue: The best-fit model spectrum assuming an atmospheric temperature of 170 K.}
\label{fig:spectrum}
\end{figure*}

For optimal placement of diagnostic SO\textsubscript{2} spectral features, observations were centered at 529.8 cm\textsuperscript{-1} (18.88 \textmu m), adjusted for Io's Doppler shift on each night. At this wavenumber, the resolving power of TEXES is $\sim$75,000 and the bandpass is $\sim$1.6 cm\textsuperscript{-1} across three cross-dispersed orders. The spectral set-up for the observations is shown in Figure~\ref{fig:spectrum}; the third order is centered on the strong SO\textsubscript{2} absorption feature at 530.4 cm\textsuperscript{-1}, while the other two orders also capture several weaker SO\textsubscript{2} absorption lines. By modeling the shapes of these absorption lines, we are able to constrain the density of Io's atmosphere. 

The TEXES instrument slit was oriented along the celestial north-south direction and observations were obtained by nodding Io 6'' along the slit. This nod-mode of observation allows for the removal of sky and telescope emission while remaining on-target. The instrument slit width at this spectral setting is 2'', which is always larger than the 0.8--1.3'' diameter of Io; our observations are therefore disk-averaged across an entire hemisphere, with a central longitude that depends on the observation time. Observations of Callisto, which is bright and spectrally gray at this spectral setting, were also obtained in order to act as a divisor for removal of instrumental and telluric effects.

Each observing sequence consisted of 16 nod pairs on Callisto ($\sim$5 mins), followed by 128 nod pairs on Io ($\sim$40 mins). When coadded, this integration time on Io provides sufficient signal-to-noise to model the SO\textsubscript{2} absorption lines while limiting the range of central longitudes we are averaging over as Io rotates. This observing sequence was typically repeated 1--3 times per night, on several nights during an observing run. This allowed us to obtain multiple independent observations of Io, each one corresponding to a different Io central longitude, as described in Table~\ref{tab:run_summary}.

\subsection{Data reduction}
\label{sec:reduction}

The nodded observations of Io and Callisto were reduced using the TEXES data reduction pipeline described in~\cite{lacy02}. This software performs flat fielding, sky subtraction, and removes the majority of instrumental geometric optical distortions. Wavelength calibration was achieved by using telluric absorption lines that fall within the spectral window. All data, including observations previously analyzed in~\cite{spencer05} and \cite{tsang12,tsang13b}, were re-reduced using the latest version of the data reduction pipeline in order to ensure consistency.

Io observations from a given observing sequence were co-added, weighted by the square of the signal-to-noise, and this spectrum was divided by the equivalent Callisto spectrum with the closest airmass. This division acts to further remove any residual instrumental effects and telluric features. Any final continuum slopes in the divided spectra were then corrected for by fitting a curve to the segments of the spectrum that are free from SO\textsubscript{2}-absorption and then dividing by this curve, and the spectrum was Doppler-shifted into Io's rest frame. Figure~\ref{fig:spectrum} shows the final reduced spectrum obtained at a longitude of 200$^{\circ}$W on 29 June 2022. Comparable spectra were produced for each of the 150 observations described in Table~\ref{tab:run_summary}. In each case, the continuum is normalized to 1, but the depths of the SO\textsubscript{2} absorption lines vary considerably due to both longitudinal and temporal variability.

\subsection{Spectral modeling}
\label{sec:spectral_modeling}

In order to retrieve the equatorial SO\textsubscript{2} column densities from the absorption spectra, we used the same method previously used by~\cite{spencer05} and~\cite{tsang12}. Because Io's atmosphere has such low density, it is not in local thermodynamic equilibrium (LTE) and must be modeled using a non-LTE model. We thus use a non-LTE model to calculate the population of the first excited state of SO\textsubscript{2} in its \textnu\textsubscript{2} mode of vibration, which is responsible for the absorption lines present at 529--531 cm\textsuperscript{-1}. We then use this model to generate synthetic disk-integrated spectra, assuming a latitudinal distribution of SO\textsubscript{2} densities based on~\cite{jessup04} and surface thermal emission based on~\cite{veeder94}. The forward model is described in greater detail in~\cite{spencer05} and~\cite{tsang12}.

This simple model has some limitations. First, in order to reduce each observation independently, we assume constant SO\textsubscript{2} column density with longitude when determining the disk-integrated model spectrum that we fit to the data~\citep{spencer05}, even though the data show that column density in fact varies with longitude.  Similarly, when determining SO\textsubscript{2} column density from frost temperature, the equatorial frost temperature is assumed to equal the diurnal peak daytime temperature at all times of day on the sunlit hemisphere.  These approximations are good to first order, given that the disk-integrated spectrum is dominated by the region within tens of degrees of the sub-Earth or sub-solar point.  However the simplifications probably somewhat blur the inferred longitudinal variability (Figure~\ref{fig:lon_fit}), and will over-estimate disk-integrated column density for a given combination of frost albedo and thermal inertia.

As described in~\cite{tsang12}, these synthetic spectra were generated for a range of equatorial SO\textsubscript{2} column densities and atmospheric temperatures and were smoothed to match the spectral resolution of the TEXES observations. A \textchi\textsuperscript{2}-minimization code was then used to find the model spectra that provided the best fit to each of the measured TEXES spectra.~\cite{tsang12} initially fitted both the atmospheric temperature and SO\textsubscript{2} column density simultaneously, but found that holding the atmospheric temperature fixed at the mean fitted value of 108 K did not significantly affect the quality of the fits. We follow the same approach here, holding the atmospheric temperature fixed at 108 K and and only allowing the SO\textsubscript{2} column density to vary when fitting each spectrum. The final best-fit model spectrum for the data obtained on 29 June 2022 at 200$^{\circ}$W is shown in red in Figure~\ref{fig:spectrum} and corresponds to an SO\textsubscript{2} column density of 1.72$\times$10\textsuperscript{17} cm\textsuperscript{-2}.

It should be noted that an atmospheric temperature of 108 K is lower than the temperatures obtained from mm-wave and near-infrared, which range from 150 K to 320 K~\citep{moullet10,roth20,depater20,lellouch15}. However, these higher atmospheric temperatures are unable to reproduce the SO\textsubscript{2} absorption lineshapes observed in the TEXES mid-infrared observations, which is a discrepancy that is not yet understood~\citep{depater23}. The blue line in Figure~\ref{fig:spectrum} shows the best-fit model spectrum when an atmospheric temperature of 170 K is used and demonstrates that is not possible to reach the observed line depths; at lower atmospheric temperatures, the line depth can be increased by increasing the SO\textsubscript{2} density, but at these higher temperatures, further increasing the SO\textsubscript{2} density leads to emission lines being produced instead of absorption lines. For the spectrum shown in Figure~\ref{fig:spectrum}, an atmospheric temperature of 93--142 K is able to fit the spectrum with a reduced chi-square statistic,~$\chi^2_\nu$\textless1.

In order to calculate the error on each fitted SO\textsubscript{2} column density value, we considered both the error in the assumed atmospheric temperature~\citep[standard deviation of 12 K,][]{tsang12} and the noise on the observed data (the standard deviation of the residual between the data and the best-fit model spectrum, shown in gray in Figure~\ref{fig:spectrum}). A synthetic `observed' spectrum was produced by adding an appropriate level of randomly-generated observational noise to the best-fit model spectrum, and it was then fitted using the same method as the real spectrum, using a randomly perturbed atmospheric temperature instead of the mean 108-K value. This process was repeated 1000 times, and the standard deviation of these SO\textsubscript{2} column densities was taken as the error on the best-fit SO\textsubscript{2} column density.

\section{Results}
\label{sec:results}

\subsection{Longitudinal variability}

\begin{figure*}
\centering
\includegraphics[width=11cm]{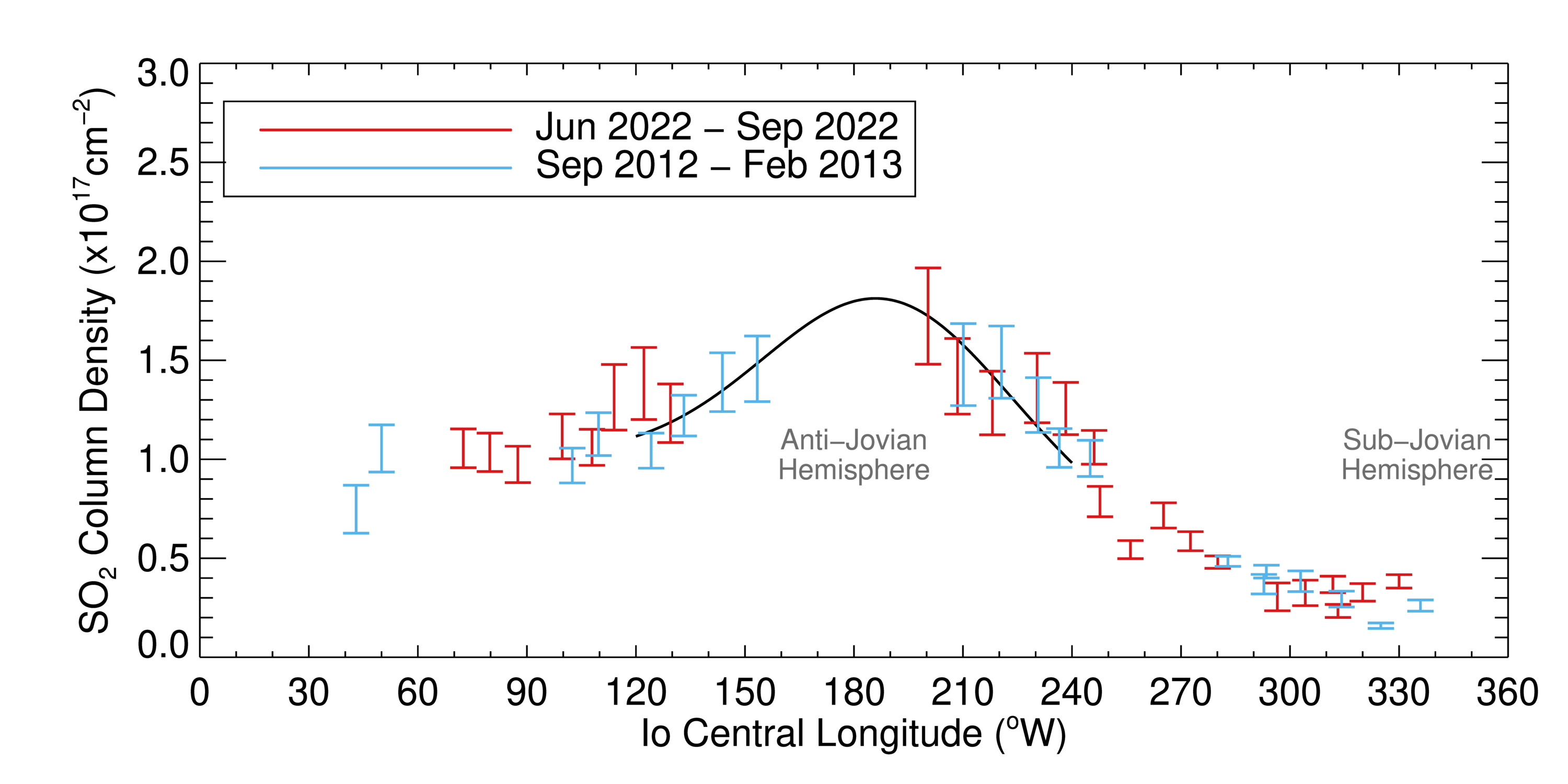}
\caption{SO\textsubscript{2} equatorial column density as a function of Io central longitude for two discrete time periods. The 2012--2013 data has been scaled by a factor of 1.1 to account for seasonally-induced temporal differences between the two time periods. The black line shows the best-fit curves for the anti-Jovian longitude region.}
\label{fig:lon_fit}
\end{figure*}

\cite{spencer05} showed that there was significant variability in the equatorial SO\textsubscript{2} column density as a function of longitude, while \cite{tsang12} and \cite{tsang13b} have shown that there is also significant temporal variability on the timescale of a Jovian year. In order to examine the longitudinal variability alone, we consider the results obtained from two discrete time periods: October 2012 -- February 2013 (shortly after perihelion) and June 2022 -- September 2022 (shortly before the following perihelion). The SO\textsubscript{2} column densities obtained in these time periods are shown in Figure~\ref{fig:lon_fit} as a function of Io central longitude. Each
time period is short enough on the Jovian-year timescale to neglect temporal variability.  An observed density variation between the two time periods is accounted for by applying a scaling factor of 1.1 to the 2012--2013 values, which minimizes the offset between the two datasets. The black line in Figure~\ref{fig:lon_fit} shows the best-fit curve (a combination of a Gaussian and a polynomial) for the anti-Jovian (120--240$^{\circ}$W) segment of the data. Due to the significant spatial variability within this range, we use this curve in Section~\ref{sec:temporal} to normalize the SO\textsubscript{2} column densities to a specific longitude in order to study temporal variability. 

Figure~\ref{fig:lon_fit} shows that in both of the time periods shown, the SO\textsubscript{2} equatorial column density varies significantly with longitude. The maximum density is located at the anti-Jupiter longitude, at $\sim$180$^{\circ}$W, while the minimum density is on the Jupiter-facing hemisphere, at $\sim$330$^{\circ}$W. The difference between the maximum and minimum densities is a factor of 5--8. These results are similar to the longitudinal results presented in \cite{spencer05} from TEXES data obtained in 2001--2004; they found the maximum was at $\sim$180$^{\circ}$W, the minimum was at $\sim$300$^{\circ}$W and the difference between the two was a factor of $\sim$10. Comparing the data from 2001--2004, 2012--2013 and 2022 shows that the longitudinal trend in the atmospheric density persists over a 20-year period. These results are further discussed in Section~\ref{sec:discussion}.

\subsection{Temporal variability on the anti-Jovian hemisphere}
\label{sec:temporal}

\begin{figure*}
\centering
\includegraphics[width=16cm]{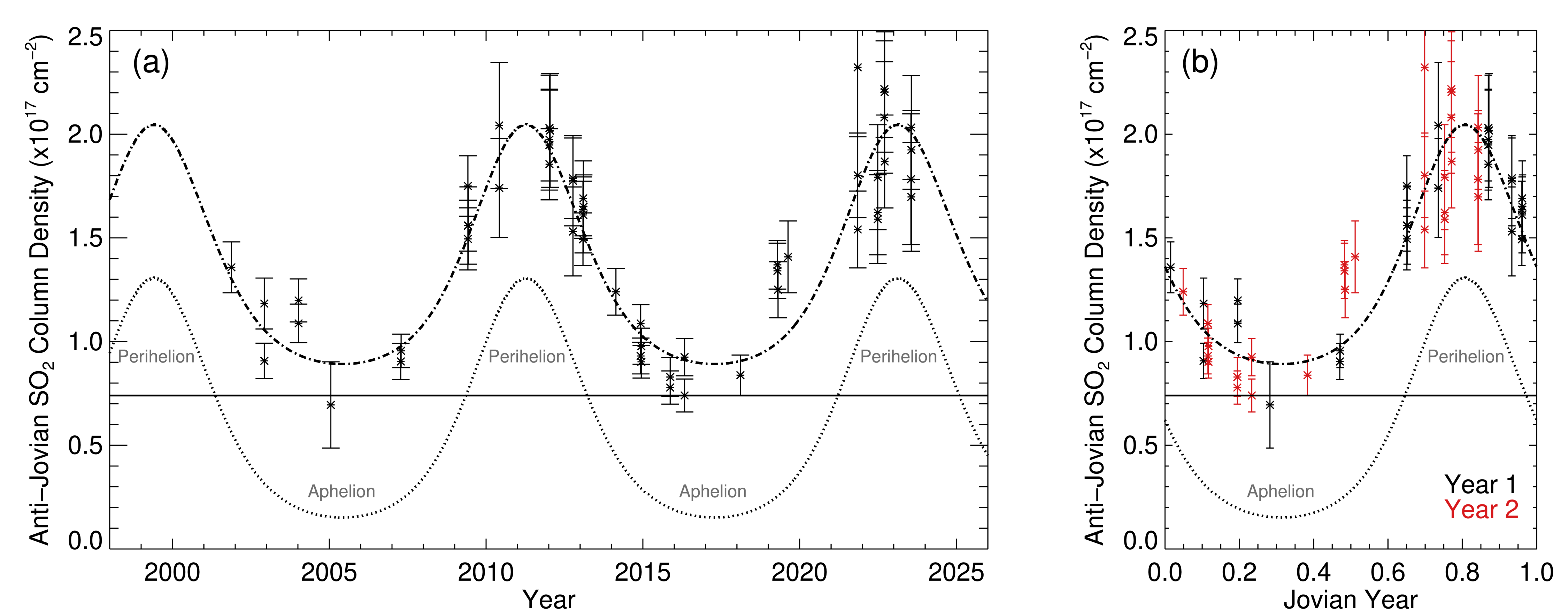}
\caption{The seasonal variability in the anti-Jovian equatorial SO\textsubscript{2} column abundances. (a) The time series from 2001--2023. (b) The same time series, but with the two Jovian years (2001-2013, and 2013-2023) superposed. The data cover longitudes of 120--240$^{\circ}$W and are normalized to 180$^{\circ}$W. The dash-dot line shows the best-fit seasonal model, combining the SO\textsubscript{2} frost vapor-pressure equilibrium curve (dotted line) and the constant, presumed volcanic, component (solid line).}
\label{fig:seasonal}
\end{figure*}

Figure~\ref{fig:seasonal} shows the long-term temporal variability in the SO\textsubscript{2} equatorial column density on the anti-Jupiter side of Io. Out of the 150 independent observations listed in Table~\ref{tab:run_summary}, Figure~\ref{fig:seasonal} presents the subset that were obtained at a central longitude of 120--240$^{\circ}$W. Figure~\ref{fig:lon_fit} shows that even within this relatively narrow longitude range, there is still considerable longitudinal variability in the SO\textsubscript{2}. In order to correct for this, we used the black fitted curve from Figure~\ref{fig:lon_fit} to normalize each measured SO\textsubscript{2} density to its equivalent 180$^{\circ}$W value. Figure~\ref{fig:seasonal}(a) shows the entire time series from 2001--2023, while Figure~\ref{fig:seasonal}(b) presents the same data, but with the two Jovian years (2001--2013 and 2013--2023) superposed. 

\begin{figure*}
\centering
\includegraphics[width=15cm]{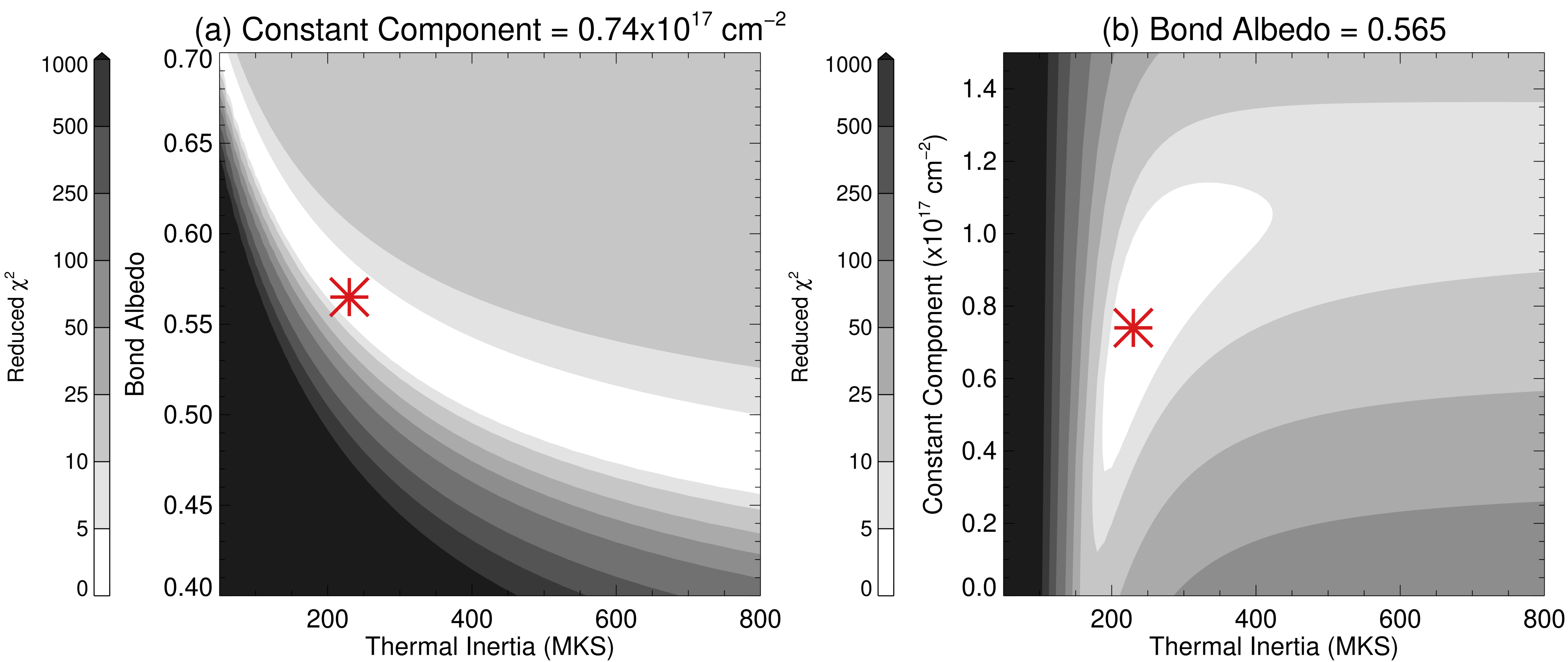}
\caption{(a) Goodness-of-fit ($\chi^2_\nu$) for the seasonal atmospheric density model fit to the anti-Jovian hemisphere data, as a function of thermal inertia and Bond albedo, for a constant component of 0.74$\times$10\textsuperscript{17} cm\textsuperscript{-2}. (b) The equivalent plot as a function of thermal inertia and constant component, for a Bond albedo of 0.565. The locations of the minimum $\chi^2_\nu$ value are marked by the red asterisks.}
\label{fig:chi2}
\end{figure*}

\cite{tsang12,tsang13b} used the TEXES data from 2001--2013 to demonstrate that there was seasonal variability in the SO\textsubscript{2} density on Io's anti-Jovian hemisphere, based on the distance of the Jovian system from the sun; they found that the Io's atmospheric density reached a minimum close to aphelion in April 2005, reached a maximum close to perihelion in March 2011 and then began to decrease again. Figure~\ref{fig:seasonal} extends the time series from \cite{tsang12,tsang13b} by another 10 years, providing almost two Jovian years of coverage. Figure~\ref{fig:seasonal}(a) shows that the SO\textsubscript{2} densities continue to follow a clear seasonal trend, with another minimum occurring around aphelion in February 2017, followed by a maximum at perihelion in January 2023. Figure~\ref{fig:seasonal}(b) compares the results from the two Jovian years in order to look for any interannual variability, but there is no apparent difference between the two time periods. At perihelion, the anti-Jovian SO\textsubscript{2} column abundance is $\sim$2.1$\times$10\textsuperscript{17} cm\textsuperscript{-2} and at aphelion, the anti-Jovian SO\textsubscript{2} column abundance is $\sim$0.9$\times$10\textsuperscript{17} cm\textsuperscript{-2}.

We modeled the data using the same seasonal model used in the original~\cite{tsang12} analysis. The dash-dot line in Figure~\ref{fig:seasonal} shows the best-fit model, which combines a seasonally-varying SO\textsubscript{2} frost vapor-pressure equilibrium component (dotted line) and a constant, presumed volcanic, component (solid line). In order to calculate the seasonally-varying component, we note that the SO\textsubscript{2} column density of Io's atmosphere is proportional to the SO\textsubscript{2} frost vapor-pressure, $P$, which in turn depends on the frost surface temperature, $T$, according to the sublimation equation from~\cite{wagman79}:

\begin{equation}
\label{eq:vaporpressure}
    P\textrm{(bar)}= 1.516 \times 10^{8} e^{-4510/T\textrm{(K)}}
\end{equation}

The surface temperature depends on both the thermophysical properties of the frost (the bolometric Bond albedo and the thermal inertia) and the distance between Jupiter and the sun, which causes the frost vapor-pressure and therefore the SO\textsubscript{2} column density to vary with season. The exponential relationship described in Equation~\ref{eq:vaporpressure} means that a modest difference in surface temperature can lead to a large change in SO\textsubscript{2} column density.

The frost surface temperatures were modeled using a one-dimension numerical thermal model that is able to model both diurnal and seasonal temperature variations~\citep{spencer89,howett11}. The model assumes that thermophysical properties are constant with depth and that sunlight is absorbed at the exact surface. The model results vary as a function of Io longitude; for the Jupiter-facing hemisphere, the model includes the effects of the $\sim$2-hr eclipse experienced during each rotation and the absorption of thermal radiation from Jupiter. 

As in \cite{tsang12,tsang13b}, this numerical model was used to calculate the frost surface temperature as a function of time for a range of Bond albedo and thermal inertia values, and these temperatures were then converted into equivalent SO\textsubscript{2} column densities as a function of time. This time-varying component was then added to a constant, assumed volcanic, component, which also covered a range of values. For all of the combinations of these three different parameters, we calculated the reduced chi-square statistic ($\chi^2_\nu$) between the seasonal model and the measured anti-Jovian SO\textsubscript{2} abundances in order to quantify the goodness-of-fit. The $\chi^2_\nu$-minimized best-fit model, shown in Figure~\ref{fig:seasonal}, corresponds to a bolometric Bond albedo of 0.565, a thermal inertia of 230 J m\textsuperscript{-2} s\textsuperscript{-1/2} K\textsuperscript{-1} (henceforth ``MKS'') and a constant, presumed volcanic, component of 0.74$\times$10\textsuperscript{17} cm\textsuperscript{-2} and has a $\chi^2_\nu$ value of 1.4. Figure~\ref{fig:chi2} shows two slices through the three-dimensional $\chi^2_\nu$ array; Figure~\ref{fig:chi2}(a) shows how $\chi^2_\nu$ varies as a function of thermal inertia and Bond albedo for the best-fit constant component of 0.74$\times$10\textsuperscript{17} cm\textsuperscript{-2}, while Figure~\ref{fig:chi2}(b) shows how $\chi^2_\nu$ varies as a function of thermal inertia and constant component, assuming the best-fit Bond albedo of 0.565. In both cases, the location of the minimum $\chi^2_\nu$ value is marked by the red asterisk, but it is clear from the shape of the contours that there is considerable degeneracy between the three parameters, in particular between the thermal inertia and the Bond albedo, and a variety of combinations of these parameters are capable of providing a comparable fit to the data. 

In order to understand the range of parameters that are able to provide a reasonable fit to the temporal data, we used a similar approach to calculating the SO\textsubscript{2} column density errors in Section~\ref{sec:spectral_modeling}. We produced a synthetic version of the temporal dataset by taking the best-fit seasonal model and adding randomly-generated noise. The noise on each datapoint is based on the observed error bars shown in Figure~\ref{fig:seasonal}, but due to the minimum $\chi^2_\nu$ value being greater than 1, we note that these error bars are likely too small and we therefore scale them up by a factor of $\sqrt{\min{(\chi^2_\nu)}}$=1.2. Using the same method as the observed temporal dataset, we then obtained the best-fit Bond albedo, thermal inertia and constant component by minimizing $\chi^2_\nu$. This process was repeated 1000 times and we calculate the median value of each parameter along with the 1-\textsigma~error. Using this method, we find that on the anti-Jovian hemisphere, the bolometric Bond albedo is 0.56$^{+0.04}_{-0.03}$, the thermal inertia is 250$^{+100}_{-90}$ MKS and the constant component is 0.74$^{+0.09}_{-0.11}\times$10\textsuperscript{17} cm\textsuperscript{-2}. Using these thermophysical frost properties, the maximum equatorial frost surface temperature on the anti-Jovian hemisphere is 116.5 K, occurring one month after perihelion and the minimum equatorial frost surface temperature on the anti-Jovian hemisphere is 110.4 K, occurring two months after aphelion.  These results are very similar to those obtained by~\cite{tsang12} using only the first Jovian year of IRTF/TEXES data.

\subsection{Temporal variability on the sub-Jovian hemisphere}
\label{sec:temporal_jovian}

\begin{figure*}
\centering
\includegraphics[width=10.2cm]{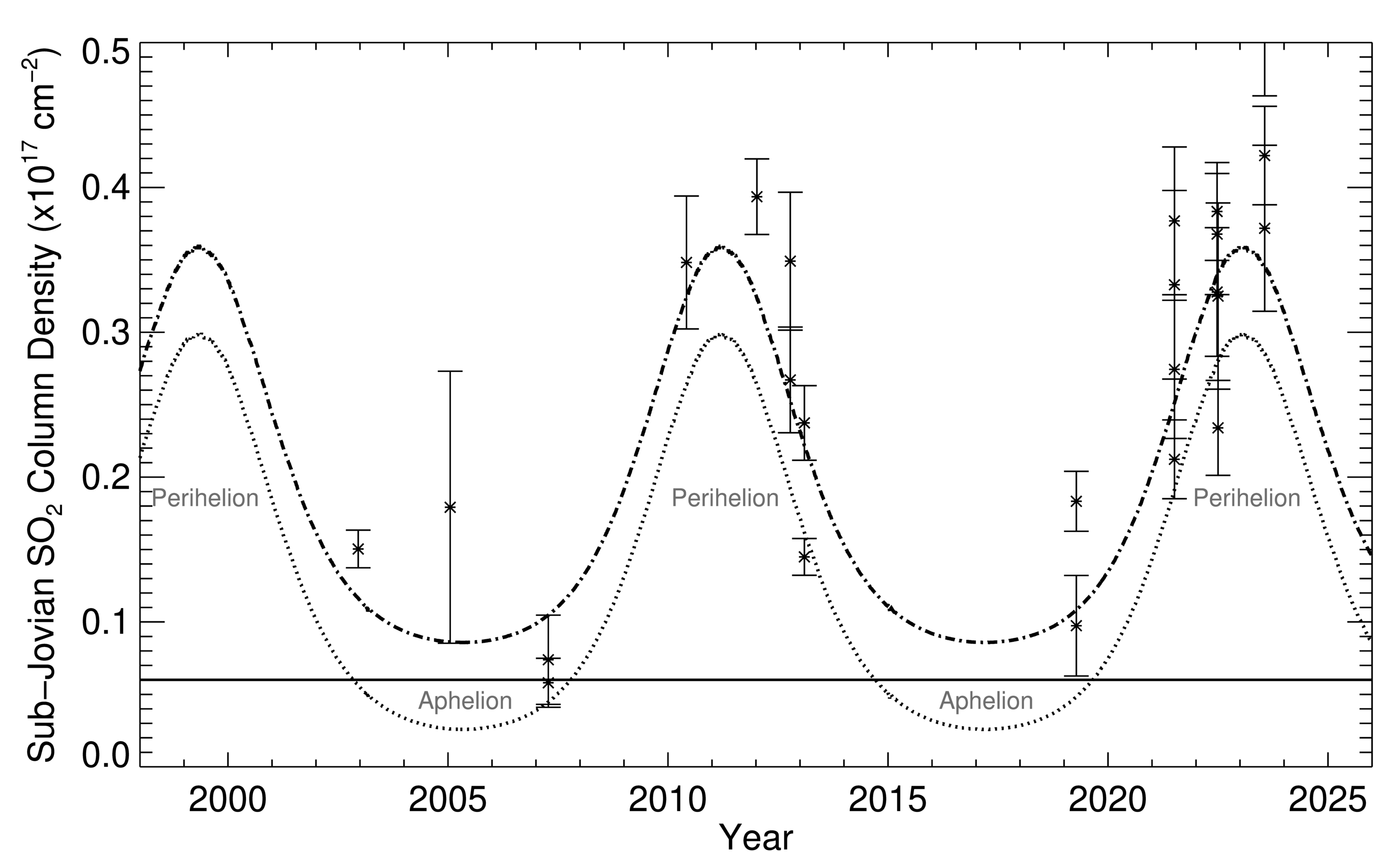}
\caption{The seasonal variability in the SO\textsubscript{2} column abundances on Io's Jupiter-facing hemisphere from 2001--2023. The data cover longitudes of 300--340$^{\circ}$W. The dash-dot line shows the best-fit seasonal model, combining the SO\textsubscript{2} frost vapor-pressure equilibrium curve (dotted line) and the constant, presumed volcanic, component (solid line).}
\label{fig:seasonal_jovian}
\end{figure*}

Figure~\ref{fig:seasonal_jovian} shows, for the first time, the SO\textsubscript{2} temporal variability on the Jupiter-facing hemisphere of Io. The data shown corresponds to observations obtained at a central longitude of 300--340$^{\circ}$W; Figure~\ref{fig:lon_fit} showed that the SO\textsubscript{2} column density is approximately constant within this range so no normalization factor was applied. Although the temporal coverage is more limited than on the anti-Jovian hemisphere, Figure~\ref{fig:seasonal_jovian} shows that the SO\textsubscript{2} column abundance on the Jupiter-facing hemisphere also follows a seasonal trend. The average densities are much lower than on the anti-Jovian hemisphere, but the temporal trend is otherwise very similar: the maximum observed SO\textsubscript{2} densities occur in 2010--2012 and 2022--2023, which correspond to times close to perihelion, while the minimum densities occur in 2003--2007 and 2019, close to aphelion. 

\begin{figure*}
\centering
\includegraphics[width=15cm]{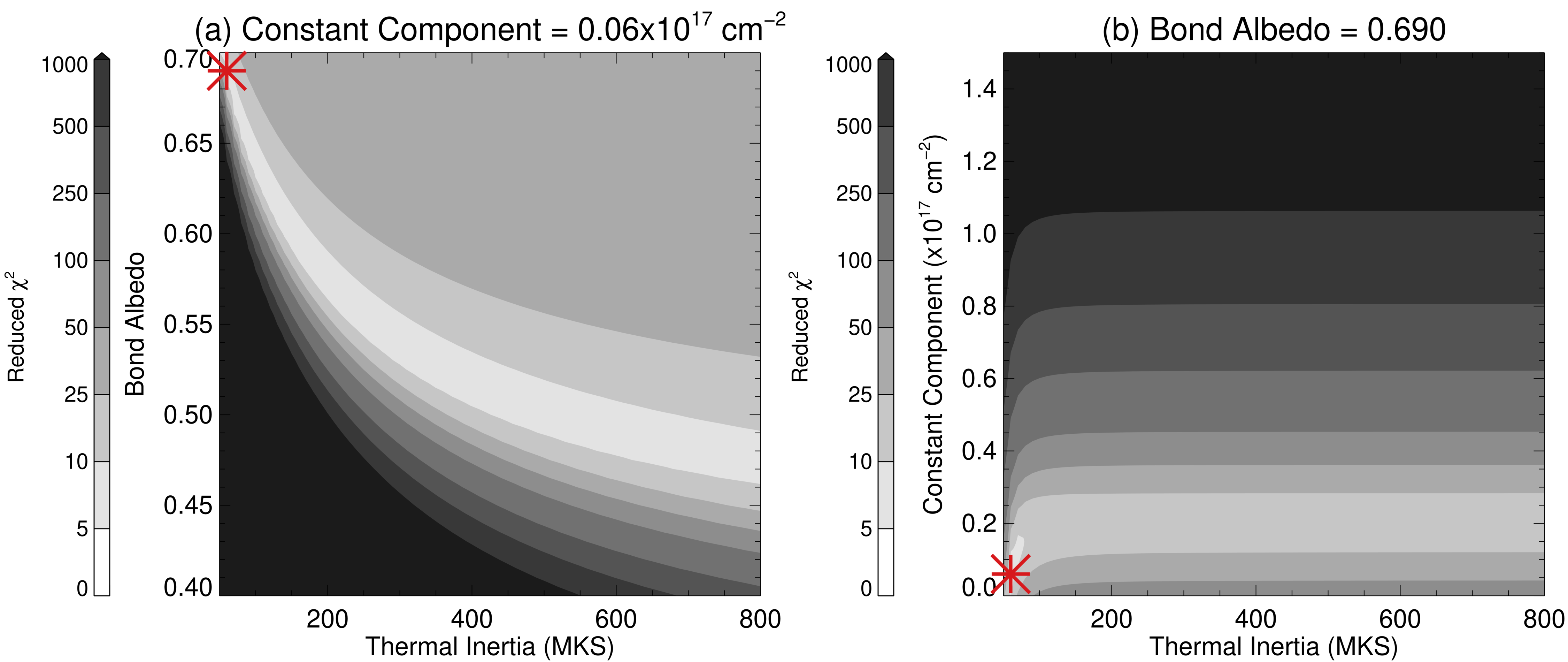}
\caption{(a) Goodness-of-fit ($\chi^2_\nu$) for the seasonal atmospheric density model fit to the sub-Jovian hemisphere data, as a function of thermal inertia and Bond albedo, for a constant component of 0.06$\times$10\textsuperscript{17} cm\textsuperscript{-2}. (b) The equivalent plot as a function of thermal inertia and constant component, for a Bond albedo of 0.690. The locations of the minimum $\chi^2_\nu$ value are marked by the red asterisks.}
\label{fig:chi2_jovian}
\end{figure*}

The black dash-dot line in Figure~\ref{fig:seasonal_jovian}(a) shows the best-fit seasonal model for the Jupiter-facing hemisphere. As in Figure~\ref{fig:seasonal}, this seasonal model consists of both a seasonally varying sublimation-driven component (dotted black line) and a constant, presumed volcanic, component (solid black line). As noted in Section~\ref{sec:temporal}, the sublimation curve for the Jupiter-facing side takes into account the effect of both eclipses and thermal radiation from Jupiter, using an effective temperature of 124.4 K for Jupiter~\citep{hanel81} and a thermal infrared albedo of 0.0. This best-fit seasonal model corresponds to a bolometric Bond albedo of 0.690, a thermal inertia of 60 MKS and a constant, presumed volcanic, component of 0.06$\times$10\textsuperscript{17} cm\textsuperscript{-2} and has a $\chi^2_\nu$ value of 5.0.

As with the anti-Jovian hemisphere, there is a considerable range of model parameters that can provide an equivalent quality of fit to the temporal data. This is demonstrated by Figure~\ref{fig:chi2_jovian}, which shows two slices through the three-dimensional $\chi^2_\nu$ array for the sub-Jovian hemisphere. As in Figure~\ref{fig:chi2}, the red asterisks show the location of the minimum $\chi^2_\nu$ value. Comparing Figures~\ref{fig:chi2}(a) and~\ref{fig:chi2_jovian}(a) shows that there is a very similar covariance between thermal inertia and Bond albedo on the two hemispheres. To formally calculate the range of acceptable values, we follow the same method used in Section~\ref{sec:temporal} where we re-run the $\chi^2_\nu$-minimization procedure for 1000 perturbed temporal datasets, with the randomly-generated noise being scaled by a factor of $\sqrt{\min{(\chi^2_\nu)}}$=2.2 from the observed error bars shown in Figure~\ref{fig:seasonal_jovian}. We find that for the Jupiter-facing hemisphere, the bolometric Bond albedo is 0.66$^{+0.03}_{-0.16}$, the thermal inertia is 80$^{+420}_{-20}$ MKS and the constant component is 0.06$^{+0.02}_{-0.06}\times$10\textsuperscript{17} cm\textsuperscript{-2}. We note that the minimum-$\chi^2_\nu$ values represented by the red asterisks in Figure~\ref{fig:chi2_jovian} are at the edge of these 1-\textsigma~error bars; the small perturbations to the data that we add during the error analysis generally produce higher thermal inertias and lower Bond albedos, which are more in line with the results obtained for the anti-Jovian hemisphere. 

The range of thermal inertia and Bond albedo values obtained from the formal error analysis shows that the same thermophysical frost properties can be used to fit both hemispheres. This is possible despite the significantly different average densities due to the effect of eclipses, which act to decrease the sublimation-driven SO\textsubscript{2} abundances. Using the same thermophysical properties obtained in Section~\ref{sec:temporal}, our model produces equatorial frost surface temperatures ranging from 106.4 to 111.9 K on the Jupiter-facing hemisphere, compared to 110.4 K to 116.5 K on the anti-Jovian hemisphere; for a given date, the sublimation-driven SO\textsubscript{2} abundances are a factor of $\sim$5 lower at a longitude of 360$^{\circ}$ than at a longitude of 180$^{\circ}$. Further information about the modeled frost surface temperatures and SO\textsubscript{2} abundances as function of season, local time and longitude can be found in~\cite{tsang12}, particularly their Figure 18.

Although the same frost thermophysical properties can be used to fit both the anti-Jovian and Jupiter-facing hemispheres, the Jupiter-facing hemisphere requires a smaller constant component, and can even be fit without a constant component altogether. The constant component on the anti-Jovian side was 0.74$^{+0.09}_{-0.11}\times$10\textsuperscript{17} cm\textsuperscript{-2}, which is higher than even the perihelion SO\textsubscript{2} densities on the Jupiter-facing side. On the Jupiter-facing hemisphere, the constant component is 0.06$^{+0.02}_{-0.06}\times$10\textsuperscript{17} cm\textsuperscript{-2}, a factor of \textgreater10 lower.

\section{Discussion and Conclusions}
\label{sec:discussion}

In this paper, we present measurements of Io's SO\textsubscript{2} atmospheric density over a period of 22 years, extending observations previously shown in~\cite{spencer05} and~\cite{tsang12,tsang13b}. The time series of SO\textsubscript{2} column abundances on Io's anti-Jovian hemisphere, where the atmosphere is thickest, now covers almost two Jovian years and shows a clear and repeated dependence on heliocentric distance, with equatorial maxima of 2.1$\times$10\textsuperscript{17} cm\textsuperscript{-2} near perihelion and minima of 0.9$\times$10\textsuperscript{17} cm\textsuperscript{-2} near aphelion. Observations from the second Jovian year follow the same trend as the first, with no apparent interannual variability.

This seasonal pattern can be well-fit using a simple model that combines a time-varying frost-sublimation curve and a constant component. The constant component could be due to direct volcanic emission; while volcanic activity does show large temporal variations on timescales of months or less it could plausibly remain approximately constant on multi-year timescales. The fact that residuals between the observations and our model in Figure~\ref{fig:seasonal} rarely exceed 0.3$\times$10\textsuperscript{17} cm\textsuperscript{-2} suggests that the discrete, short-lived volcanic events seen in ground-based surveys~\citep[e.g.][]{dekleer17,tate23} rarely contribute more than this value to the disk-integrated atmospheric column density.

At perihelion, when the surface temperatures are highest, our model suggests that the sublimation component on the anti-Jupiter hemisphere provides 64$\pm$5\% of the atmospheric density on the anti-Jovian hemisphere while the constant component provides 36$\pm$5\%. At aphelion, when the surface temperatures are lowest, the dominant mechanism is reversed, with the constant component providing 82$\pm$11\% and the sublimation component providing 18$\pm$11\%. The recent observations continue to support the conclusions made by~\cite{tsang12,tsang13b} that both frost sublimation and direct volcanic emissions provide a significant contribution to Io's atmosphere.  However, while the consistent seasonal variability over two years provides unambiguous evidence for the importance of sublimation support, the limitations of our seasonal atmospheric model, noted above, mean that caution is required when inferring the properties of the  constant, potentially volcanic, component. In our model this constant component is required to reduce the amplitude of the model seasonal variation to match the observed amplitude, but it is possible that a more sophisticated model might be able to match the observed amplitude using sublimation support alone. The long-term stability of the atmosphere demonstrated by this paper (after accounting for longitudinal and seasonal effects), and by Ly-\textalpha~imaging \citep{giono21}, is expected from sublimation support, given the >40 year stability of the surface SO\textsubscript{2} frost distribution that is evident in visible-wavelength imaging from Voyager \citep{mcewen88} to New Horizons~\citep{spencer07}, to Juno~\citep{mura23}.

On the anti-Jovian hemisphere, we obtain a bolometric Bond albedo of 0.56$^{+0.04}_{-0.03}$ and a thermal inertia of 250$^{+100}_{-90}$ MKS. These results are in agreement with the previous value obtained by~\cite{tsang12} but have smaller error bars due to the additional observations that have been made over the past ten years. These thermophysical frost properties are also completely consistent with previous independent measurements. Using Galileo images of Io, \cite{simonelli01} found that the average Bond albedo over the surface was 0.51$\pm$0.09. This value combines regions with and without SO\textsubscript{2} ice, so in order to constrain the albedo of SO\textsubscript{2} ice alone, \cite{tsang12} compared the maps of Bond albedo from \cite{simonelli01} with maps of fractional frost coverage from \cite{doute01} to estimate that the Bond albedo of SO\textsubscript{2} ice on Io's surface is no less than 0.55. \cite{walker12} used this constraint, along with a combination of atmospheric column density observations from HST/STIS ~\citep{jessup04} and surface temperature measurements from Galileo's Photo-Polarimeter Radiometer (PPR)~\citep{rathbun04} to infer that the Bond albedo of the SO\textsubscript{2} frost was 0.55$\pm$0.02 and the thermal inertia was 200$\pm$50 MKS. 

The SO\textsubscript{2} atmospheric density results that we present in this paper show the long-term persistence of the previously reported large longitudinal variability, with the maximum density at $\sim$180$^{\circ}$ being 5--8 times higher than the minimum density at $\sim$330$^{\circ}$. Previous results from TEXES observations in 2001--2004, found a factor of $\sim$10 difference between the maximum at $\sim$180$^{\circ}$W and the minimum at $\sim$300$^{\circ}$W~\citep{spencer05}. Ly-\textalpha~images from HST/STIS obtained over multiple years also show higher SO\textsubscript{2} densities on the anti-Jovian hemisphere~\citep{feaga09,giono21}, although it should be noted that the error bars on these results are large~\citep{giono21}.

The cause of this longitudinal variability in the SO\textsubscript{2} density is not yet well established. If the difference is driven by sublimation, it could be due to spatial differences in the SO\textsubscript{2} surface ice thermophysical properties or fractional coverage, which varies significantly with longitude~\citep{mcewen88} in a way that closely matches the atmospheric distribution \citep{spencer05}. In simple vapor pressure equilibrium models, surface fractional frost coverage will not affect column density, but such simple models may not apply to Io's dynamic, patchy atmosphere, for instance if residence times for SO\textsubscript{2} molecules in the non-SO\textsubscript{2} surface components are long~\citep{walker10}. Even if there is no variation in the surface ice, the Jupiter-facing hemisphere experiences regular eclipses by Jupiter that will act to lower the diurnally-averaged surface temperatures relative to the anti-Jovian side, decreasing the sublimation of the surface ice and therefore decreasing the atmospheric density. The additional thermal radiation from Jupiter impinging on the Jupiter-facing hemisphere is not sufficient to compensate for the loss of sunlight during eclipses. The eclipse effect was proposed by \cite{tsang12} and \cite{walker12} as a driver of the longitudinal variability. The \cite{walker12} atmospheric simulation results showed the sub-Jupiter point having an average dayside atmospheric density that is 4 times lower than the anti-Jupiter point, not very different from the observed ratio of 5--8. Our numerical thermal model produces a similar factor of 4--5 decrease in the sublimation-driven SO\textsubscript{2} density. Another contribution to the longitudinal variability could be variations in direct volcanic supply, given the inhomogeneous spatial distribution of active plumes, which are more abundant on the anti-Jovian hemisphere~\citep{mcewen83,spencer07,geissler08}.

Although the Jupiter-facing data are more sparse and are noisier due to the weaker absorption lines, our IRTF/TEXES observations show that the seasonal dependence that is clearly observed on the anti-Jovian hemisphere is also present on the Jupiter-facing hemisphere. This shows that sublimation must play an important role in atmospheric support on both hemispheres, consistent with the Jupiter eclipse studies discussed in Section~\ref{sec:introduction}. On the Jupiter-facing hemisphere, the SO\textsubscript{2} column density reaches 0.4$\times$10\textsuperscript{17} cm\textsuperscript{-2} near perihelion and drops to 0.1$\times$10\textsuperscript{17} cm\textsuperscript{-2} near aphelion. We find that we are able to fit this sub-Jovian seasonal pattern using the same thermophysical ice parameters as the anti-Jovian hemisphere; while different albedo and thermal inertia values are certainly possible, such differences are not required by the data and the lower sublimation-driven atmospheric densities can be plausibly explained by the daily eclipses. The very low atmospheric densities seen near aphelion show that any constant volcanic component on the sub-Jovian hemisphere must be small, which may also contribute to the spatial inhomogeneity in the atmospheric density. 

\section*{Acknowledgements}

The authors were Visiting Astronomers at the Infrared Telescope Facility, which is operated by the University of Hawaii under contract 80HQTR19D0030 with the National Aeronautics and Space Administration. This work was funded by NASA Solar System Observations Grant 80NSSC21K1138. 

\printcredits

\bibliographystyle{cas-model2-names}

\begin{thebibliography}{34}
\expandafter\ifx\csname natexlab\endcsname\relax\def\natexlab#1{#1}\fi
\providecommand{\url}[1]{\texttt{#1}}
\providecommand{\href}[2]{#2}
\providecommand{\path}[1]{#1}
\providecommand{\DOIprefix}{doi:}
\providecommand{\ArXivprefix}{arXiv:}
\providecommand{\URLprefix}{URL: }
\providecommand{\Pubmedprefix}{pmid:}
\providecommand{\doi}[1]{\href{http://dx.doi.org/#1}{\path{#1}}}
\providecommand{\Pubmed}[1]{\href{pmid:#1}{\path{#1}}}
\providecommand{\bibinfo}[2]{#2}
\ifx\xfnm\relax \def\xfnm[#1]{\unskip,\space#1}\fi
%Type = Article
\bibitem[{De~Pater et~al.(2020)De~Pater, Luszcz-Cook, Rojo, Redwing, De~Kleer
  and Moullet}]{depater20}
\bibinfo{author}{De~Pater, I.}, \bibinfo{author}{Luszcz-Cook, S.},
  \bibinfo{author}{Rojo, P.}, \bibinfo{author}{Redwing, E.},
  \bibinfo{author}{De~Kleer, K.}, \bibinfo{author}{Moullet, A.},
  \bibinfo{year}{2020}.
\newblock \bibinfo{title}{{ALMA} observations of {Io} going into and coming out
  of eclipse}.
\newblock \bibinfo{journal}{The Planetary Science Journal} \bibinfo{volume}{1},
  \bibinfo{pages}{60}.
%Type = Article
\bibitem[{Dout{\'e} et~al.(2001)Dout{\'e}, Schmitt, Lopes-Gautier, Carlson,
  Soderblom, Shirley, Team et~al.}]{doute01}
\bibinfo{author}{Dout{\'e}, S.}, \bibinfo{author}{Schmitt, B.},
  \bibinfo{author}{Lopes-Gautier, R.}, \bibinfo{author}{Carlson, R.},
  \bibinfo{author}{Soderblom, L.}, \bibinfo{author}{Shirley, J.},
  \bibinfo{author}{Team, G.N.}, et~al., \bibinfo{year}{2001}.
\newblock \bibinfo{title}{Mapping {SO\textsubscript{2}} frost on {Io} by the
  modeling of {NIMS} hyperspectral images}.
\newblock \bibinfo{journal}{Icarus} \bibinfo{volume}{149},
  \bibinfo{pages}{107--132}.
%Type = Article
\bibitem[{Feaga et~al.(2009)Feaga, McGrath and Feldman}]{feaga09}
\bibinfo{author}{Feaga, L.M.}, \bibinfo{author}{McGrath, M.},
  \bibinfo{author}{Feldman, P.D.}, \bibinfo{year}{2009}.
\newblock \bibinfo{title}{Io's dayside {SO\textsubscript{2}} atmosphere}.
\newblock \bibinfo{journal}{Icarus} \bibinfo{volume}{201},
  \bibinfo{pages}{570--584}.
%Type = Article
\bibitem[{Geissler and McMillan(2008)}]{geissler08}
\bibinfo{author}{Geissler, P.E.}, \bibinfo{author}{McMillan, M.T.},
  \bibinfo{year}{2008}.
\newblock \bibinfo{title}{Galileo observations of volcanic plumes on {Io}}.
\newblock \bibinfo{journal}{Icarus} \bibinfo{volume}{197},
  \bibinfo{pages}{505--518}.
%Type = Article
\bibitem[{Giono and Roth(2021)}]{giono21}
\bibinfo{author}{Giono, G.}, \bibinfo{author}{Roth, L.}, \bibinfo{year}{2021}.
\newblock \bibinfo{title}{Io's {SO\textsubscript{2}} atmosphere from {HST
  Lyman-$\alpha$} images: 1997 to 2018}.
\newblock \bibinfo{journal}{Icarus} \bibinfo{volume}{359},
  \bibinfo{pages}{114212}.
%Type = Article
\bibitem[{Hanel et~al.(1981)Hanel, Conrath, Herath, Kunde and
  Pirraglia}]{hanel81}
\bibinfo{author}{Hanel, R.A.}, \bibinfo{author}{Conrath, B.J.},
  \bibinfo{author}{Herath, L.W.}, \bibinfo{author}{Kunde, V.G.},
  \bibinfo{author}{Pirraglia, J.A.}, \bibinfo{year}{1981}.
\newblock \bibinfo{title}{Albedo, internal heat, and energy balance of
  {Jupiter}: Preliminary results of the {Voyager} infrared investigation}.
\newblock \bibinfo{journal}{Journal of Geophysical Research: Space Physics}
  \bibinfo{volume}{86}, \bibinfo{pages}{8705--8712}.
%Type = Article
\bibitem[{Howett et~al.(2011)Howett, Spencer, Pearl and Segura}]{howett11}
\bibinfo{author}{Howett, C.J.A.}, \bibinfo{author}{Spencer, J.R.},
  \bibinfo{author}{Pearl, J.}, \bibinfo{author}{Segura, M.},
  \bibinfo{year}{2011}.
\newblock \bibinfo{title}{High heat flow from {Enceladus'} south polar region
  measured using 10--600 cm\textsuperscript{-1} {Cassini/CIRS} data}.
\newblock \bibinfo{journal}{Journal of Geophysical Research: Planets}
  \bibinfo{volume}{116}.
%Type = Article
\bibitem[{Jessup and Spencer(2015)}]{jessup15}
\bibinfo{author}{Jessup, K.L.}, \bibinfo{author}{Spencer, J.R.},
  \bibinfo{year}{2015}.
\newblock \bibinfo{title}{Spatially resolved {HST/STIS} observations of {Io's}
  dayside equatorial atmosphere}.
\newblock \bibinfo{journal}{Icarus} \bibinfo{volume}{248},
  \bibinfo{pages}{165--189}.
%Type = Article
\bibitem[{Jessup et~al.(2004)Jessup, Spencer, Ballester, Howell, Roesler, Vigel
  and Yelle}]{jessup04}
\bibinfo{author}{Jessup, K.L.}, \bibinfo{author}{Spencer, J.R.},
  \bibinfo{author}{Ballester, G.E.}, \bibinfo{author}{Howell, R.R.},
  \bibinfo{author}{Roesler, F.}, \bibinfo{author}{Vigel, M.},
  \bibinfo{author}{Yelle, R.}, \bibinfo{year}{2004}.
\newblock \bibinfo{title}{The atmospheric signature of {Io's Prometheus} plume
  and anti-jovian hemisphere: evidence for a sublimation atmosphere}.
\newblock \bibinfo{journal}{Icarus} \bibinfo{volume}{169},
  \bibinfo{pages}{197--215}.
%Type = Article
\bibitem[{de~Kleer and de~Pater(2017)}]{dekleer17}
\bibinfo{author}{de~Kleer, K.}, \bibinfo{author}{de~Pater, I.},
  \bibinfo{year}{2017}.
\newblock \bibinfo{title}{{Io's Loki Patera}: Modeling of three brightening
  events in 2013--2016}.
\newblock \bibinfo{journal}{Icarus} \bibinfo{volume}{289},
  \bibinfo{pages}{181--198}.
%Type = Article
\bibitem[{Lacy et~al.(2002)Lacy, Richter, Greathouse, Jaffe and Zhu}]{lacy02}
\bibinfo{author}{Lacy, J.H.}, \bibinfo{author}{Richter, M.J.},
  \bibinfo{author}{Greathouse, T.K.}, \bibinfo{author}{Jaffe, D.T.},
  \bibinfo{author}{Zhu, Q.}, \bibinfo{year}{2002}.
\newblock \bibinfo{title}{{TEXES}: A sensitive high-resolution grating
  spectrograph for the mid-infrared}.
\newblock \bibinfo{journal}{Publications of the Astronomical Society of the
  Pacific} \bibinfo{volume}{114}, \bibinfo{pages}{153--168}.
%Type = Article
\bibitem[{Lellouch et~al.(2015)Lellouch, Ali-Dib, Jessup, Smette, K{\"a}ufl and
  Marchis}]{lellouch15}
\bibinfo{author}{Lellouch, E.}, \bibinfo{author}{Ali-Dib, M.},
  \bibinfo{author}{Jessup, K.L.}, \bibinfo{author}{Smette, A.},
  \bibinfo{author}{K{\"a}ufl, H.U.}, \bibinfo{author}{Marchis, F.},
  \bibinfo{year}{2015}.
\newblock \bibinfo{title}{Detection and characterization of {Io's} atmosphere
  from high-resolution 4-$\mu$m spectroscopy}.
\newblock \bibinfo{journal}{Icarus} \bibinfo{volume}{253},
  \bibinfo{pages}{99--114}.
%Type = Article
\bibitem[{McEwen et~al.(1988)McEwen, Johnson, Matson and Soderblom}]{mcewen88}
\bibinfo{author}{McEwen, A.S.}, \bibinfo{author}{Johnson, T.V.},
  \bibinfo{author}{Matson, D.L.}, \bibinfo{author}{Soderblom, L.A.},
  \bibinfo{year}{1988}.
\newblock \bibinfo{title}{The global distribution, abundance, and stability of
  {SO\textsubscript{2}} on {Io}}.
\newblock \bibinfo{journal}{Icarus} \bibinfo{volume}{75},
  \bibinfo{pages}{450--478}.
%Type = Article
\bibitem[{McEwen and Soderblom(1983)}]{mcewen83}
\bibinfo{author}{McEwen, A.S.}, \bibinfo{author}{Soderblom, L.A.},
  \bibinfo{year}{1983}.
\newblock \bibinfo{title}{Two classes of volcanic plumes on {Io}}.
\newblock \bibinfo{journal}{Icarus} \bibinfo{volume}{55},
  \bibinfo{pages}{191--217}.
%Type = Article
\bibitem[{Moullet et~al.(2010)Moullet, Gurwell, Lellouch and
  Moreno}]{moullet10}
\bibinfo{author}{Moullet, A.}, \bibinfo{author}{Gurwell, M.A.},
  \bibinfo{author}{Lellouch, E.}, \bibinfo{author}{Moreno, R.},
  \bibinfo{year}{2010}.
\newblock \bibinfo{title}{Simultaneous mapping of {SO\textsubscript{2}, SO,
  NaCl} in {Io's} atmosphere with the {Submillimeter Array}}.
\newblock \bibinfo{journal}{Icarus} \bibinfo{volume}{208},
  \bibinfo{pages}{353--365}.
%Type = Inproceedings
\bibitem[{Mura et~al.(2023)Mura, Tosi, Zamboni, Hansen, Lopes, Becker, Rathbun,
  Adriani, Plainaki, Sindoni et~al.}]{mura23}
\bibinfo{author}{Mura, A.}, \bibinfo{author}{Tosi, F.},
  \bibinfo{author}{Zamboni, F.}, \bibinfo{author}{Hansen, C.},
  \bibinfo{author}{Lopes, R.M.}, \bibinfo{author}{Becker, H.N.},
  \bibinfo{author}{Rathbun, J.}, \bibinfo{author}{Adriani, A.},
  \bibinfo{author}{Plainaki, C.}, \bibinfo{author}{Sindoni, G.}, et~al.,
  \bibinfo{year}{2023}.
\newblock \bibinfo{title}{Juno observations of {Io}}, in:
  \bibinfo{booktitle}{EGU Abstracts}.
%Type = Incollection
\bibitem[{de~Pater et~al.(2023)de~Pater, Goldstein and Lellouch}]{depater23}
\bibinfo{author}{de~Pater, I.}, \bibinfo{author}{Goldstein, D.},
  \bibinfo{author}{Lellouch, E.}, \bibinfo{year}{2023}.
\newblock \bibinfo{title}{{The Plumes and Atmosphere of Io}}, in:
  \bibinfo{booktitle}{{Io: A New View of Jupiter's Moon}}.
  \bibinfo{publisher}{Springer}. chapter \bibinfo{chapter}{{The Plumes and
  Atmosphere of Io}}, pp. \bibinfo{pages}{233--290}.
%Type = Article
\bibitem[{Rathbun et~al.(2004)Rathbun, Spencer, Tamppari, Martin, Barnard and
  Travis}]{rathbun04}
\bibinfo{author}{Rathbun, J.A.}, \bibinfo{author}{Spencer, J.R.},
  \bibinfo{author}{Tamppari, L.K.}, \bibinfo{author}{Martin, T.Z.},
  \bibinfo{author}{Barnard, L.}, \bibinfo{author}{Travis, L.D.},
  \bibinfo{year}{2004}.
\newblock \bibinfo{title}{Mapping of {Io's} thermal radiation by the {Galileo}
  photopolarimeter--radiometer {(PPR)} instrument}.
\newblock \bibinfo{journal}{Icarus} \bibinfo{volume}{169},
  \bibinfo{pages}{127--139}.
%Type = Article
\bibitem[{Retherford et~al.(2007)Retherford, Spencer, Stern, Saur, Strobel,
  Steffl, Gladstone, Weaver, Cheng, Parker et~al.}]{retherford07}
\bibinfo{author}{Retherford, K.}, \bibinfo{author}{Spencer, J.},
  \bibinfo{author}{Stern, S.}, \bibinfo{author}{Saur, J.},
  \bibinfo{author}{Strobel, D.}, \bibinfo{author}{Steffl, A.},
  \bibinfo{author}{Gladstone, G.}, \bibinfo{author}{Weaver, H.},
  \bibinfo{author}{Cheng, A.}, \bibinfo{author}{Parker, J.W.}, et~al.,
  \bibinfo{year}{2007}.
\newblock \bibinfo{title}{Io's atmospheric response to eclipse: {UV} aurorae
  observations}.
\newblock \bibinfo{journal}{Science} \bibinfo{volume}{318},
  \bibinfo{pages}{237--240}.
%Type = Article
\bibitem[{Roth et~al.(2020)Roth, Boissier, Moullet, S{\'a}nchez-Monge,
  de~Kleer, Yoneda, Hikida, Kita, Tsuchiya, Bl{\"o}cker et~al.}]{roth20}
\bibinfo{author}{Roth, L.}, \bibinfo{author}{Boissier, J.},
  \bibinfo{author}{Moullet, A.}, \bibinfo{author}{S{\'a}nchez-Monge, {\'A}.},
  \bibinfo{author}{de~Kleer, K.}, \bibinfo{author}{Yoneda, M.},
  \bibinfo{author}{Hikida, R.}, \bibinfo{author}{Kita, H.},
  \bibinfo{author}{Tsuchiya, F.}, \bibinfo{author}{Bl{\"o}cker, A.}, et~al.,
  \bibinfo{year}{2020}.
\newblock \bibinfo{title}{An attempt to detect transient changes in {Io's}
  {SO\textsubscript{2} and NaCl} atmosphere}.
\newblock \bibinfo{journal}{Icarus} \bibinfo{volume}{350},
  \bibinfo{pages}{113925}.
%Type = Article
\bibitem[{Saur and Strobel(2004)}]{saur04}
\bibinfo{author}{Saur, J.}, \bibinfo{author}{Strobel, D.F.},
  \bibinfo{year}{2004}.
\newblock \bibinfo{title}{Relative contributions of sublimation and volcanoes
  to {Io's} atmosphere inferred from its plasma interaction during solar
  eclipse}.
\newblock \bibinfo{journal}{Icarus} \bibinfo{volume}{171},
  \bibinfo{pages}{411--420}.
%Type = Article
\bibitem[{Simonelli et~al.(2001)Simonelli, Dodd and Veverka}]{simonelli01}
\bibinfo{author}{Simonelli, D.P.}, \bibinfo{author}{Dodd, C.},
  \bibinfo{author}{Veverka, J.}, \bibinfo{year}{2001}.
\newblock \bibinfo{title}{Regolith variations on {Io}: Implications for
  bolometric albedos}.
\newblock \bibinfo{journal}{Journal of Geophysical Research: Planets}
  \bibinfo{volume}{106}, \bibinfo{pages}{33241--33252}.
%Type = Article
\bibitem[{Spencer et~al.(1989)Spencer, Lebofsky and Sykes}]{spencer89}
\bibinfo{author}{Spencer, J.R.}, \bibinfo{author}{Lebofsky, L.A.},
  \bibinfo{author}{Sykes, M.V.}, \bibinfo{year}{1989}.
\newblock \bibinfo{title}{Systematic biases in radiometric diameter
  determinations}.
\newblock \bibinfo{journal}{Icarus} \bibinfo{volume}{78},
  \bibinfo{pages}{337--354}.
%Type = Article
\bibitem[{Spencer et~al.(2005)Spencer, Lellouch, Richter, L{\'o}pez-Valverde,
  Jessup, Greathouse and Flaud}]{spencer05}
\bibinfo{author}{Spencer, J.R.}, \bibinfo{author}{Lellouch, E.},
  \bibinfo{author}{Richter, M.J.}, \bibinfo{author}{L{\'o}pez-Valverde, M.A.},
  \bibinfo{author}{Jessup, K.L.}, \bibinfo{author}{Greathouse, T.K.},
  \bibinfo{author}{Flaud, J.M.}, \bibinfo{year}{2005}.
\newblock \bibinfo{title}{Mid-infrared detection of large longitudinal
  asymmetries in {Io's SO\textsubscript{2}} atmosphere}.
\newblock \bibinfo{journal}{Icarus} \bibinfo{volume}{176},
  \bibinfo{pages}{283--304}.
%Type = Article
\bibitem[{Spencer et~al.(2007)Spencer, Stern, Cheng, Weaver, Reuter,
  Retherford, Lunsford, Moore, Abramov, Lopes et~al.}]{spencer07}
\bibinfo{author}{Spencer, J.R.}, \bibinfo{author}{Stern, S.A.},
  \bibinfo{author}{Cheng, A.F.}, \bibinfo{author}{Weaver, H.A.},
  \bibinfo{author}{Reuter, D.C.}, \bibinfo{author}{Retherford, K.},
  \bibinfo{author}{Lunsford, A.}, \bibinfo{author}{Moore, J.M.},
  \bibinfo{author}{Abramov, O.}, \bibinfo{author}{Lopes, R.M.C.}, et~al.,
  \bibinfo{year}{2007}.
\newblock \bibinfo{title}{Io volcanism seen by {New Horizons}: A major eruption
  of the {Tvashtar} volcano}.
\newblock \bibinfo{journal}{Science} \bibinfo{volume}{318},
  \bibinfo{pages}{240--243}.
%Type = Article
\bibitem[{Tate et~al.(2023)Tate, Rathbun, Hayes, Spencer and Pettine}]{tate23}
\bibinfo{author}{Tate, C.D.}, \bibinfo{author}{Rathbun, J.A.},
  \bibinfo{author}{Hayes, A.G.}, \bibinfo{author}{Spencer, J.R.},
  \bibinfo{author}{Pettine, M.}, \bibinfo{year}{2023}.
\newblock \bibinfo{title}{Discovery of seven volcanic outbursts on {Io} from an
  {Infrared Telescope Facility} observation campaign, 2016--2022}.
\newblock \bibinfo{journal}{The Planetary Science Journal} \bibinfo{volume}{4},
  \bibinfo{pages}{189}.
%Type = Article
\bibitem[{Tsang et~al.(2013a)Tsang, Spencer and Jessup}]{tsang13}
\bibinfo{author}{Tsang, C.C.}, \bibinfo{author}{Spencer, J.R.},
  \bibinfo{author}{Jessup, K.L.}, \bibinfo{year}{2013}a.
\newblock \bibinfo{title}{Synergistic observations of {Io's atmosphere in 2010
  from HST--COS in the mid-ultraviolet and IRTF--TEXES in the mid-infrared}}.
\newblock \bibinfo{journal}{Icarus} \bibinfo{volume}{226},
  \bibinfo{pages}{604--616}.
%Type = Article
\bibitem[{Tsang et~al.(2016)Tsang, Spencer, Lellouch, Lopez-Valverde and
  Richter}]{tsang16}
\bibinfo{author}{Tsang, C.C.}, \bibinfo{author}{Spencer, J.R.},
  \bibinfo{author}{Lellouch, E.}, \bibinfo{author}{Lopez-Valverde, M.A.},
  \bibinfo{author}{Richter, M.J.}, \bibinfo{year}{2016}.
\newblock \bibinfo{title}{The collapse of {Io's} primary atmosphere in
  {Jupiter} eclipse}.
\newblock \bibinfo{journal}{Journal of Geophysical Research: Planets}
  \bibinfo{volume}{121}, \bibinfo{pages}{1400--1410}.
%Type = Article
\bibitem[{Tsang et~al.(2012)Tsang, Spencer, Lellouch, L{\'o}pez-Valverde,
  Richter and Greathouse}]{tsang12}
\bibinfo{author}{Tsang, C.C.C.}, \bibinfo{author}{Spencer, J.R.},
  \bibinfo{author}{Lellouch, E.}, \bibinfo{author}{L{\'o}pez-Valverde, M.A.},
  \bibinfo{author}{Richter, M.J.}, \bibinfo{author}{Greathouse, T.K.},
  \bibinfo{year}{2012}.
\newblock \bibinfo{title}{Io's atmosphere: Constraints on sublimation support
  from density variations on seasonal timescales using {NASA IRTF/TEXES}
  observations from 2001 to 2010}.
\newblock \bibinfo{journal}{Icarus} \bibinfo{volume}{217},
  \bibinfo{pages}{277--296}.
%Type = Article
\bibitem[{Tsang et~al.(2013b)Tsang, Spencer, Lellouch, L{\'o}pez-Valverde,
  Richter, Greathouse and Roe}]{tsang13b}
\bibinfo{author}{Tsang, C.C.C.}, \bibinfo{author}{Spencer, J.R.},
  \bibinfo{author}{Lellouch, E.}, \bibinfo{author}{L{\'o}pez-Valverde, M.A.},
  \bibinfo{author}{Richter, M.J.}, \bibinfo{author}{Greathouse, T.K.},
  \bibinfo{author}{Roe, H.}, \bibinfo{year}{2013}b.
\newblock \bibinfo{title}{Io's contracting atmosphere post 2011 perihelion:
  further evidence for partial sublimation support on the anti-{Jupiter}
  hemisphere}.
\newblock \bibinfo{journal}{Icarus} \bibinfo{volume}{226},
  \bibinfo{pages}{1177--1181}.
%Type = Article
\bibitem[{Veeder et~al.(1994)Veeder, Matson, Johnson, Blaney and
  Goguen}]{veeder94}
\bibinfo{author}{Veeder, G.J.}, \bibinfo{author}{Matson, D.L.},
  \bibinfo{author}{Johnson, T.V.}, \bibinfo{author}{Blaney, D.L.},
  \bibinfo{author}{Goguen, J.D.}, \bibinfo{year}{1994}.
\newblock \bibinfo{title}{{Io's} heat flow from infrared radiometry:
  1983--1993}.
\newblock \bibinfo{journal}{Journal of Geophysical Research: Planets}
  \bibinfo{volume}{99}, \bibinfo{pages}{17095--17162}.
%Type = Article
\bibitem[{Wagman(1979)}]{wagman79}
\bibinfo{author}{Wagman, D.D.}, \bibinfo{year}{1979}.
\newblock \bibinfo{title}{Sublimation pressure and enthalpy of
  {SO\textsubscript{2}}}.
\newblock \bibinfo{journal}{{Chem. Thermodynamics Data Center Natl. Bureau
  Standards, Washington, DC}} .
%Type = Article
\bibitem[{Walker et~al.(2010)Walker, Gratiy, Goldstein, Moore, Varghese,
  Trafton, Levin and Stewart}]{walker10}
\bibinfo{author}{Walker, A.C.}, \bibinfo{author}{Gratiy, S.L.},
  \bibinfo{author}{Goldstein, D.B.}, \bibinfo{author}{Moore, C.H.},
  \bibinfo{author}{Varghese, P.L.}, \bibinfo{author}{Trafton, L.M.},
  \bibinfo{author}{Levin, D.A.}, \bibinfo{author}{Stewart, B.},
  \bibinfo{year}{2010}.
\newblock \bibinfo{title}{A comprehensive numerical simulation of {Io's}
  sublimation-driven atmosphere}.
\newblock \bibinfo{journal}{Icarus} \bibinfo{volume}{207},
  \bibinfo{pages}{409--432}.
%Type = Article
\bibitem[{Walker et~al.(2012)Walker, Moore, Goldstein, Varghese and
  Trafton}]{walker12}
\bibinfo{author}{Walker, A.C.}, \bibinfo{author}{Moore, C.H.},
  \bibinfo{author}{Goldstein, D.B.}, \bibinfo{author}{Varghese, P.L.},
  \bibinfo{author}{Trafton, L.M.}, \bibinfo{year}{2012}.
\newblock \bibinfo{title}{A parametric study of {Io's} thermophysical surface
  properties and subsequent numerical atmospheric simulations based on the best
  fit parameters}.
\newblock \bibinfo{journal}{Icarus} \bibinfo{volume}{220},
  \bibinfo{pages}{225--253}.

\end{thebibliography}


\end{document}